\documentclass[prl, twocolumn, floatfix]{revtex4}
\setcitestyle{super}

\usepackage{graphicx}
\usepackage{hyperref}
\usepackage{amsmath}
\usepackage{color}
\usepackage[normalem]{ulem}
\usepackage[abs]{overpic}

\newcommand{\doi}[2]{\href{http://dx.doi.org/#1}{#2}}

\begin{document}

\title{A surface-patterned chip as a strong source of ultra-cold atoms for quantum technologies}
\author{C.\ C.\ Nshii,$^{1}$\footnote{These authors contributed equally.} M.\ Vangeleyn,$^{1*}$ J.\ P.\ Cotter,$^{2*}$ P.\ F.\ Griffin,$^1$ E.\ A.\ Hinds,$^2$
C.\ N.\ Ironside,$^3$ P.\ See,$^4$ A.\ G.\ Sinclair,$^4$ E.\ Riis,$^1$ and A.\ S.\ Arnold$^1$} \affiliation{$^{1}$Dept.\ of Physics, SUPA,  University of Strathclyde, Glasgow G4 0NG, UK}
\affiliation{$^{2}$Centre for Cold Matter, Blackett Laboratory, Imperial College London, Prince Consort Road, London SW7 2BW, UK}
\affiliation{$^{3}$Rankine Building, School of Engineering, University of Glasgow, Glasgow G12 8LT, UK}
\affiliation{$^{4}$National Physical Laboratory, Hampton Road, Teddington, Middlesex, TW11 0LW, UK}

\date{\today}

\maketitle

\textbf{Laser cooled atoms are central to modern precision measurements \cite{takamoto05,deutsch10,buning11,bodart10,poli11,lamporesi08}.  They are also increasingly important as an enabling technology for experimental cavity quantum electrodynamics\cite{specht11,ritter12}, quantum information processing\cite{bakr10,sherson10,weimer10} and matter wave interferometry\cite{muller10}. Although significant progress has been made in miniaturising atomic metrological devices\cite{knappe03,shah07}, these are limited in accuracy by their use of hot atomic ensembles and buffer gases. Advances have also been made in producing portable apparatus that benefit from the advantages of atoms in the microKelvin regime\cite{vanzoest10,atomchipbook}. However, simplifying atomic cooling and loading using microfabrication technology has proved difficult\cite{pollock09,pollock11}. 
In this letter we address this problem, realising an atom chip that enables the integration of laser cooling and trapping into a compact apparatus. Our source delivers ten thousand times more atoms than previous magneto-optical traps with microfabricated optics and, for the first time, can reach sub-Doppler temperatures. Moreover, the same chip design offers a simple way to form stable optical lattices. These features, combined with the simplicity of fabrication and the ease of operation, make these new traps a key advance in the development of cold-atom technology for high-accuracy, portable measurement devices.}

There have been rapid developments in quantum technology since the first microKelvin atom traps\cite{raab87}. Today many core experimental techniques can be realised using atom chips\cite{atomchipbook} --- microfabricated structures that trap, guide and detect ultra-cold atoms in a small integrated package. These offer the prospect of miniature, portable instruments, such as clocks and magnetometers, based on the internal quantum states of atoms. Bose-Einstein condensates can also be produced on a chip \cite{hansel01}, giving access to the quantum behaviour of the atomic motion for applications using phenomena such as matter-wave interference\cite{schumm05,baumgartner10} or spin entanglement\cite{riedel10}. Atom chips offer the tantalising prospect of portable setups, as ultra-high-vacuum chambers\cite{coldquanta} and  cooling lasers are now available in small packages. However, the main obstacle to miniaturisation has been the lack of a simple integrated magneto-optical trap (MOT) to collect and cool the atoms initially and to load them onto the chip.

A MOT is formed near the zero of a quadrupole magnetic field, located within the overlap region of four or more appropriately polarised laser beams, slightly red-detuned from an atomic resonance\cite{raab87}, see \href{http://photonics.phys.strath.ac.uk/wp-content/uploads/2013/11/NanofabGratingSuppR4.pdf}{Supplementary Section}. A pyramidal reflector can produce the required beams from a single circularly-polarised input\cite{lee96,vangeleyn09}. Such reflectors have recently been integrated into an atom chip\cite{pollock09,pollock11}, however, the number of atoms captured is less than $7 \times 10^{3}$ -- far too low for applications requiring degeneracy or where signal to noise is paramount. The atoms are also trapped below the chip surface, making them inconvenient to access and detect. Vangeleyn {\em et. al.} pointed out that the pyramid mirrors can be replaced by diffraction gratings\cite{vangeleyn10}. Here we describe a chip with diffraction gratings microfabricated directly into the surface, which forms a MOT from a single input beam. Despite its simplicity this MOT delivers $6 \times 10^{7}$ atoms in a readily accessible trap - as many as a conventional 6-beam MOT of the same volume and four orders of magnitude more than the micro-pyramid MOT\cite{pollock11}. 

\begin{figure*}[t!]
\centering
\includegraphics[width=2.08\columnwidth]{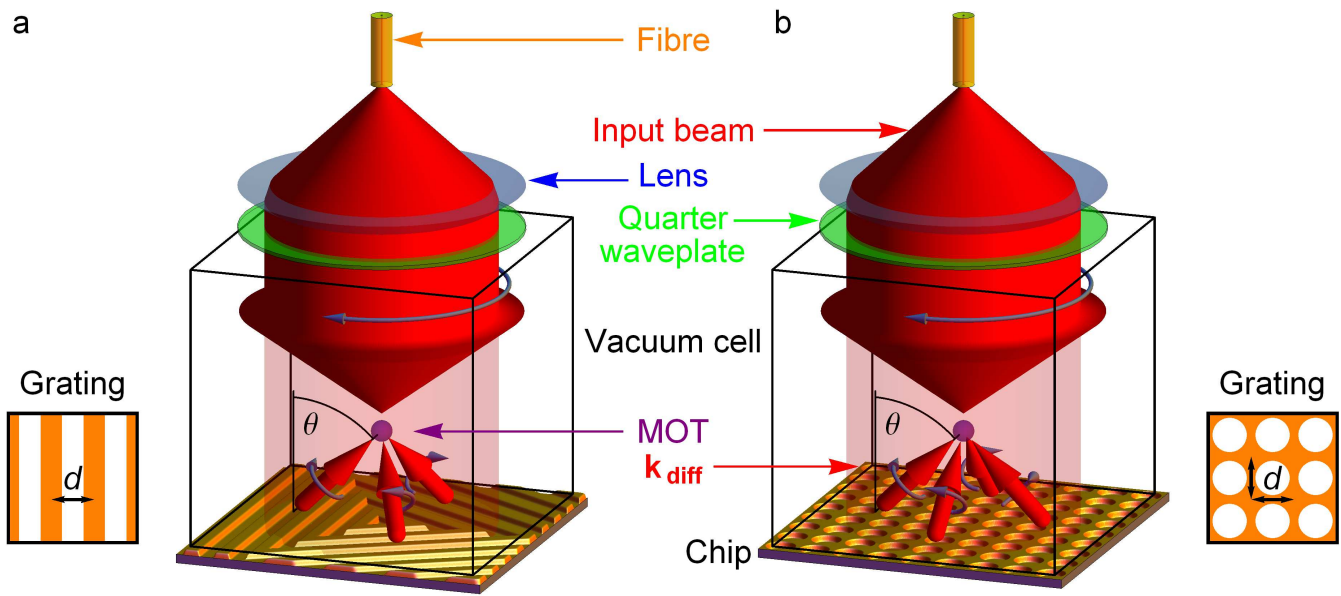}
\caption{\label{Fig1}
\textbf{\textbf{Concept of the grating chip MOT.}} Linearly-polarised light (red) diverging from the output of an optical fibre (orange) is collimated and circularly polarised (grey arrow) by the combination of a lens (blue) and quarter-wave plate (green). This single input beam diffracts from microfabricated gratings on the chip to produce the additional beams (small red arrows indicate wave-vectors, \textbf{k}$_{\rm diff}$) needed to form a MOT. The light traps atoms from a low-pressure vapour inside an evacuated glass cell, see Methods. 
\textbf{a} Three linear gratings (pattern shown inset) diffract the light into the $n_{x} = \pm 1$ orders to form a 4-beam MOT (only trapping beams are shown). \textbf{b} A square array of cylindrical indentations (pattern shown inset) diffracts the input into the $n_{x} = \pm 1$ and $n_{y} = \pm 1$ orders to form a 5-beam MOT. The retro-reflections ($n_x, n_y =0$) are strongly suppressed and higher orders of diffraction are eliminated (see Methods). Magnetic quadrupole coils are omitted for clarity.
}
\end{figure*}

A range of binary patterns can provide the appropriate laser beam configuration for a MOT. Fig.\,\ref{Fig1}a shows three linear gratings producing a tetrahedral configuration of beams, while Fig.\,\ref{Fig1}b shows a square lattice of cylindrical indentations making a five-beam arrangement. The diffraction angle $\theta$ depends on the wavelength $\lambda$ of the light and the period  $d$ of the grating through the Bragg conditions $n_{x}\lambda = d \sin{\theta}$ (linear) and  $\sqrt{{n_x}^2+{n_y}^2}\lambda = d \sin{\theta}$ (square).

\begin{figure*}[t!]
\centering
\includegraphics[width=2\columnwidth]{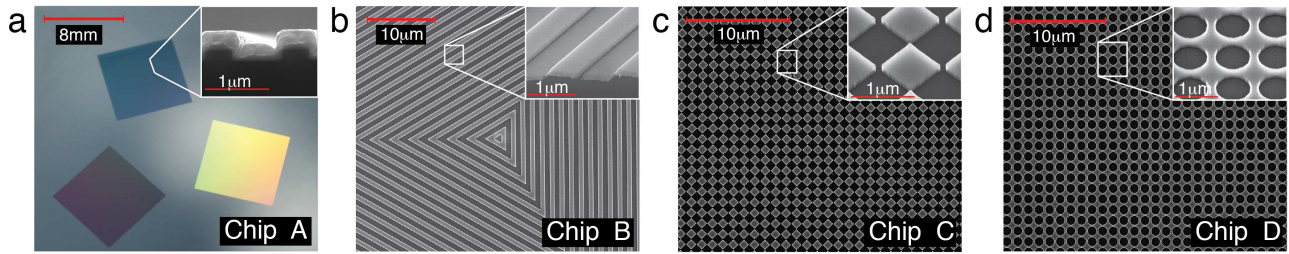}
\caption{\label{Fig2} \textbf{\textbf{The grating chips.}}  \textbf{a} Photograph of chip A which consists of three linear $8\,\mbox{mm}\times8\,\mbox{mm}$ gratings of pitch $d=1200\,$nm ($\theta =41^{\circ}$).  
Inset: scanning electron microscope (SEM) image showing the cleaved cross section through the coated grating.  \textbf{b} Chip B extends the pattern of chip A inward to the centre and outward over a $20\,\mbox{mm}\times 20\,\mbox{mm}$ area. 
The SEM image shows the centre of the mask, which has a pitch of $d=1400\,$nm  ($\theta =34^{\circ}$). Inset: SEM close-up of the etched and coated chip surface. \textbf{c} Chip C is an array of squares covering a total area of $2$\,cm$\times2$\,cm, with a pitch of $d=1080\,$nm ($\theta =46^{\circ}$). An SEM image of the mask is shown, together with a zoomed image, inset. \textbf{d} Chip D is the same as C, but with the squares replaced by circles.
}
\end{figure*}

The four particularly simple grating chip designs we investigated are presented in Fig.\,\ref{Fig2}, in order of increasing capture volume.  We use standard semiconductor processing techniques to fabricate chips in either Si or GaAs. This approach has considerable scope for scalable production, particularly using nanoimprint technology.
Chip A was produced by photolithography, whereas Chips B-D were made by electron beam lithography. Full details of the fabrication process and reflective coatings for each chip are discussed in the Methods. Normally, atom chips are placed inside an ultra-high vacuum (UHV) chamber as trapped atoms have to be protected from collisions with background gas. Our chips can be operated in a UHV chamber but they can also be used outside, with a vacuum window between the surface of the chip and the cold atom cloud.
All four chips have been tested in this simple configuration, using a single input laser beam. 

\begin{figure*}[!t]
\begin{center}$
\begin{array}{ccc}
\begin{overpic}[height=5cm]{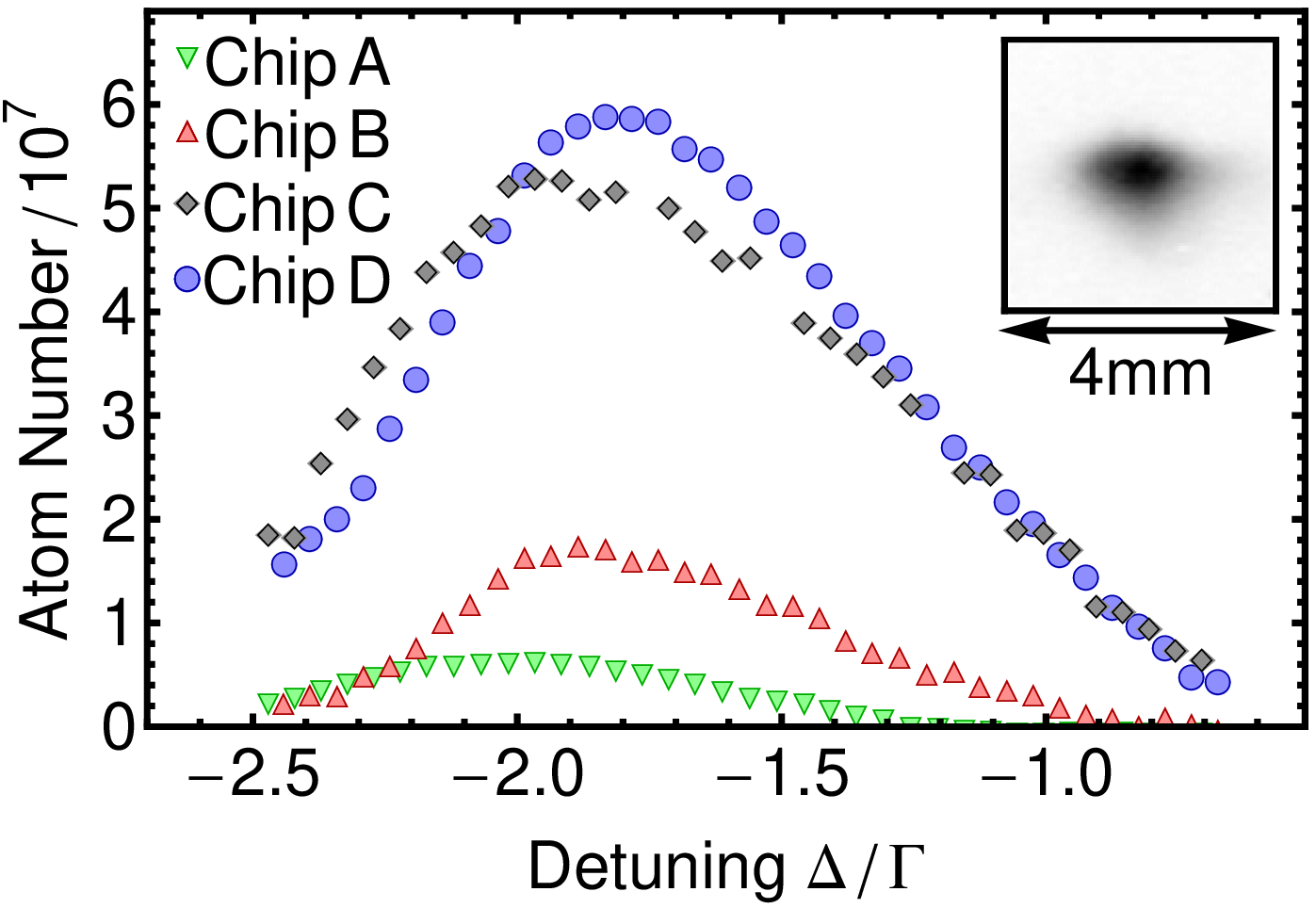}
\put(-14,130){\large \textsf{a}}
\end{overpic} 
\hspace{1cm}
\begin{overpic}[height=5cm]{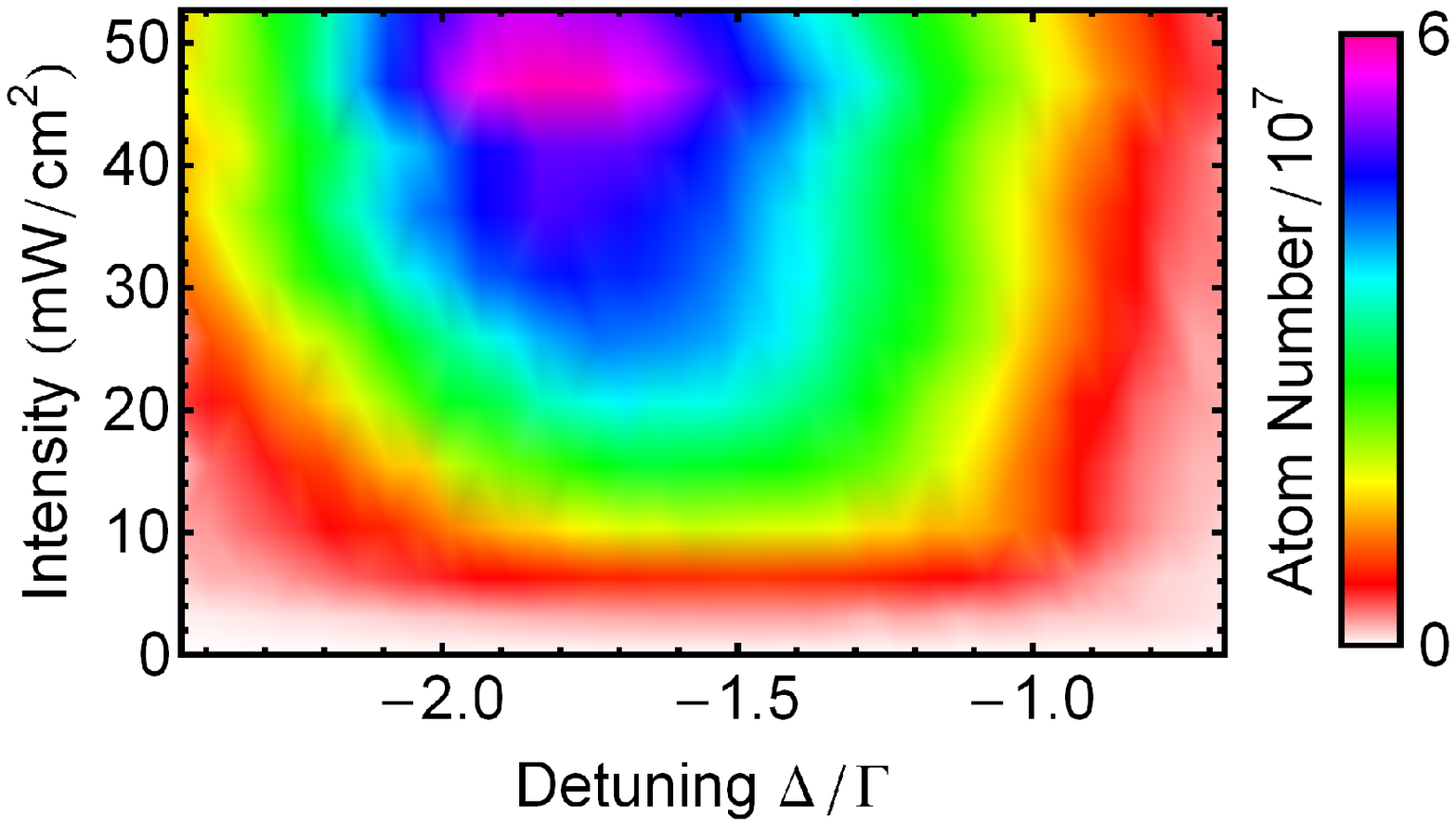}
\put(-10,130){\large \textsf{b}}
\end{overpic}
\end{array}$
\end{center}
\vspace{-5mm}
\caption{\label{fig:det} \textbf{Variation of atom number with laser detuning and intensity}. \textbf{a} Number of atoms trapped by each chip as the detuning was varied, measured in units of natural linewidth $\Gamma = 2\pi \times 6\,$MHz. For each curve a fixed intensity, which optimised the atom number, was used. Data points are the average of 5 runs, resulting in statistical uncertainties much smaller than the plot markers. The atom number peaks at a detuning around two linewidths below resonance, which is typical for most MOTs. Inset: Fluorescence of trapped atoms above chip D at maximum atom number. \textbf{b} Atom number versus detuning and input beam peak intensity using chip D, obtained using the detunings in \textbf{a} at 11 additional intensities and interpolating a surface.}
\end{figure*}

The number of atoms trapped in a MOT depends on the beam overlap volume, as well as the laser frequency and intensity. Figure\,\ref{fig:det}a shows the number of $^{87}$Rb atoms trapped by each chip from a low-pressure vapour as the laser frequency is varied. The detuning, which optimises the number of atoms trapped by each chip, is typical of conventional MOTs of the same trapping volume. 
In Fig.\,\ref{fig:det}b, we see chip D's dependence on both detuning and intensity. The behaviour shown here is representative of all the chip MOTs described in this letter and MOTs in general. Around the peak, the number of atoms depends only weakly on the intensity and frequency of the light, resulting in a stable MOT number.
The beam overlap volume determines the atom number because each trap dimension is equivalent to a stopping distance, which determines the maximum speed of atoms which can be trapped. 
Chip B collects three times as many atoms as chip A because its diffracted beams have larger cross section, producing a larger trap volume and hence capturing faster atoms \cite{lind, pollock11}. Chips C and D capture three times as many atoms again, because this geometry has a differently shaped and larger overlap volume (\href{http://photonics.phys.strath.ac.uk/wp-content/uploads/2013/11/NanofabGratingSuppR4.pdf}{Supplementary Section}).

\begin{figure}[b!]
\centering\includegraphics[width=0.85\columnwidth]{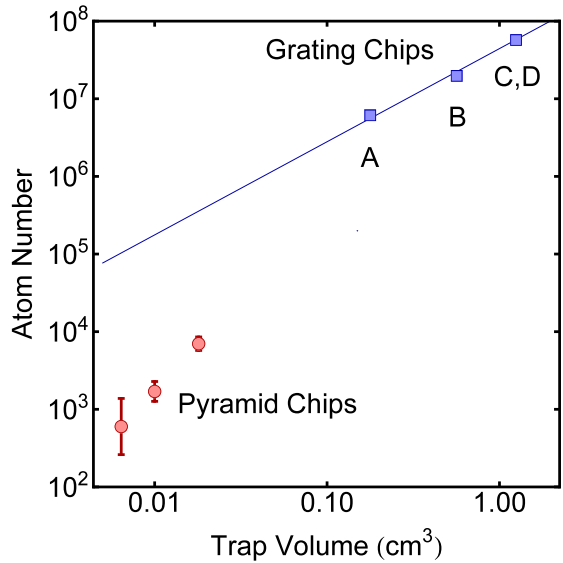}
\caption{\label{Scaling}
{\bf Variation of peak atom number, $N$,  with trapping volume, $V$.} Blue squares: number of atoms trapped by each of the grating chips A-D. Red circles: numbers trapped by microfabricated pyramid chips\cite{pollock11}. Where error bars are not visible, they are much smaller than the data points.
}
\end{figure}

In Fig.\,\ref{Scaling} we plot the peak number of trapped atoms $N$ for each chip versus the trap volume $V$.   The trap dimensions are large enough to follow the approximate scaling law\cite{lind} $N\propto V^{1.2}$, indicated by the blue line in Fig.\,\ref{Scaling}. The maximum number of atoms captured by grating Chip D is $6\times 10^{7}$, essentially the same as a conventional six-beam MOT of the same volume (see Methods). By contrast, the pyramid MOT of Pollock \textit{et al.}\cite{pollock11}, the only other MOT in the literature with microfabricated optics, follows a $V^2$ power law because of its small volume, capturing to date a maximum of $7 \times 10^{3}$ atoms as indicated in Fig.\,\ref{Scaling} by the red points.

\begin{figure}[t!]
\centering
\includegraphics[width=.9\columnwidth]{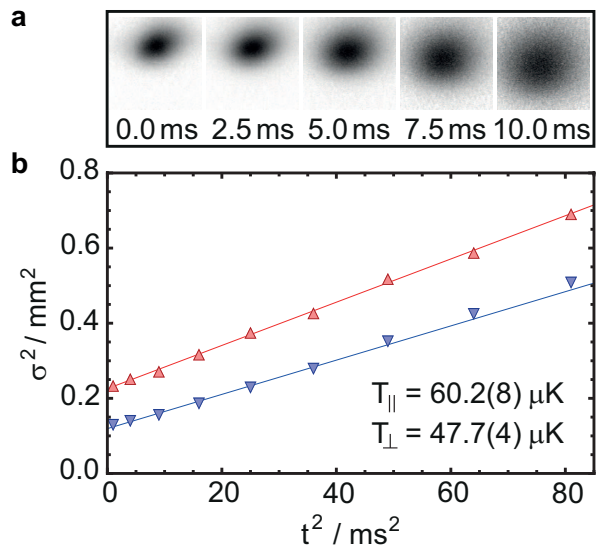}
\caption{\label{Fig5} \textbf{Temperature measurements on chip B}. \textbf{a} Sequence of fluorescence images ($2.7\,$mm$\times 2.7\,$mm) taken during ballistic expansion, after optical molasses. \textbf{b} Mean square cloud radii parallel (red upward triangles) and perpendicular (blue downward triangles) to the grating, versus square of expansion time $t$. Error bars are smaller than the size of a point. Lines: fits of $\sigma^{2} = \sigma_{0}^{2} + \frac{k_{B} T}{m}t^2$ to the data, where $k_{B}$ is the Boltzmann constant, $T$ is the temperature and $m$ is the mass of an $^{87}$Rb atom. Anisotropic expansion of a molasses is not unusual.
}
\end{figure}

For many applications, such as the production of quantum degenerate gases, sub-Doppler temperatures are important in order to obtain high phase space densities. We have demonstrated that our chips can reach this regime. The initial temperature of atoms trapped on our chips is $\sim 1\,$mK, typical of standard MOTs and a few times the Doppler temperature ($140\mu$K for Rb). The MOT temperature depends on the laser intensity and detuning and the diffraction angle. We are able to lower the temperature more than an order of magnitude, through sub-Doppler mechanisms\cite{dalibard89}. The slow expansion after optical molasses, using Chip B, is plotted in Fig.\,\ref{Fig5}, yielding a sub-Doppler temperature of $50$--$60\,\mu$K. The molasses works well for chip B because the upward and downward radiation pressures are well balanced.  
The combination of large atom number and low temperature makes the grating MOTs suitable for clocks and atom interferometry. While it does not provide the same long interaction time as an atomic fountain it offers the advantage of absolute stability at a vastly reduced size and level of complexity.

There is a fixed relative phase between the diffracted beams, referenced to the chip surface. This ensures a stable, periodic interference pattern of intensity and polarisation that requires no user alignment.
In this way, an input light beam diffracting from anywhere on grating chips C or D automatically creates a three-dimensional body-centred-cubic lattice of microscopic atom traps which, for suitable far red-detuned light, localises atoms in intensity maxima due to the optical dipole force. Optical lattices are already well established as a valuable technique in atomic clocks\cite{takamoto05} and quantum simulators\cite{blochrev}. Our gratings open the possibility of introducing lattices in a simple way to atom chips.

In conclusion we have demonstrated that optical gratings microfabricated into an atom chip can be used to produce a MOT and to make optical molasses for sub-Doppler cooling. When illuminated by a $20\,$mm laser beam, which determines the optical trapping volume, the grating MOT traps $\sim10^8$ atoms. It will be straightforward to produce a fully integrated device with magnetic field wires fabricated on the chip\cite{atomchipbook}. This would also allow multiple traps.

In comparison with the standard four-beam reflection MOT\cite{reichel99} our design has two main advantages. First, the MOT needs only one laser beam and therefore requires much less optical access. Second, a single circularly polarised input beam requires no further optics. This makes the implementation and the alignment extremely simple. We also note that each trapping beam passes only once through the atomic cloud, minimising the intensity imbalance due to absorption. In addition, the gratings offer a simple and stable way to introduce 3D optical lattices onto chips, significantly extending their functionality. This level of optical integration was previously missing and opens the way for devices built on atom chips with simpler optics and a smaller footprint.

\noindent
{\large \bf Methods}\\
{\footnotesize
\noindent
{\bf Chip fabrication.}
We etched patterns to a depth of $\lambda/4$ ($\lambda=780\,$nm), with an approximately 50:50 etched:unetched area ratio, in order to suppress zeroth-order diffraction (reflection). For all the chips, this gave less than 1\% of the incident power in the zeroth-order. The first order diffracted beams are described below for each chip. In all cases over $96\,$\% of the diffracted power had the correct polarisation for MOT operation. Higher diffraction orders are cut off because $d\le2\lambda$ for the linear gratings and $d \le \sqrt{2}\lambda$ for the square gratings.
\\
\textit{\bf Chip A:}
Fabricated in silicon using reactive ion etching (RIE). The groove depth was measured to be $210\,(10)\,$nm using an atomic force microscope (AFM). This chip was sputter coated with $200\,$nm of gold, following a $10\,$nm adhesion layer of chromium, resulting in a reflectivity of $\simeq 0.97$. We measured 40(1)\% of the incident light in each of the $\pm$ first order diffracted beams. The missing power ($\sim 20\%$) is scattered out of the first order beams because the propagating electric field is disturbed by the vertical walls of the grating trenches.
\\
\textit{\bf Chip B:}
Fabricated in silicon using RIE. The AFM measured a groove depth of $206\,(10)\,$nm. With a $3\,$nm nickel-chromium adhesion layer, and a $100\,$nm aluminium reflection coating we measured 38(1)\% of incident power in each first order diffracted beam. With the additional diffracted beam losses due to two uncoated glass surfaces in the cell, this chip is close to the ideal $1/3$ diffraction efficiency required for optimally balanced light forces in optical molasses\cite{vangeleyn10}.
\\
\textit{\bf Chips C and D:}
Fabricated in silicon or gallium-arsenide. The silicon chips had the same groove depth as chip B. For the gallium arsenide versions an etch stop was used to ensure a uniform depth of $195(5)\,$nm. The same reflective coating as B was used for both. We found very little difference in the optical behaviour of the coated Si and GaAs chips. Both varieties were tested inside and outside the vacuum chamber and had comparable performance. The data shown in this letter is from the silicon chips. We measured that $20(1)$\% of the incident power was diffracted into each of the four first order beams.\\  \\
\noindent
{\bf Vacuum, magnetic fields and imaging.}
All experiments used an externally anti-reflection coated quartz vacuum chamber with inner dimensions $22 \times 22 \times 80\,$mm$^{3}$. A base pressure $<10^{-9}\,$mbar was maintained by a $40\,$l/s ion pump. A rubidium partial pressure $<5\times10^{-9}\,$mbar was regulated by a metal dispenser. The magnetic quadrupole coils, aligned coaxially with the chip normal, were operated with an axial gradient of $10\,$G/cm.
Three orthogonal pairs of Helmholtz coils cancelled the ambient magnetic field. The MOT fluorescence was viewed using a CCD camera placed $10\,$cm from the atoms, using an imaging system with transverse magnification of 0.80.\noindent
\\ \\
{\bf Optics and laser system.} Two external cavity diode lasers were frequency stabilised on the $^{87}$Rb D2 line using saturated absorption spectroscopy. One was locked to the $F=1\rightarrow2$ hyperfine transition required for repumping. The other was locked with a red detuning of $0-33\,$MHz relative to the $F=2\rightarrow3$ hyperfine transition. The latter injected a slave laser that produced up to $20\,$mW of cooling light. This was combined with the repumping light on a beam splitter, linearly co-polarised and coupled into a single-mode optical fibre. A fibre-coupled tapered amplifier provided optical gain up to a total power of $500\,$mW. Its fibre output, shown in Fig.\,\ref{Fig1}, was collimated to form a Gaussian beam of $20\,$mm $1/e^2$ intensity radius, and circularly polarised using a quarter-wave plate immediately outside the vacuum window. This beam was adjusted to be at normal incidence to the grating chip outside the opposite window. The MOT was switched off in less than $10\, \mu$s using an acousto-optic modulator to deflect the beam injecting the slave laser, restoring the laser to its natural frequency 1\,THz to the red of the cooling transition. This has the same effect on the atoms as switching off the light, and has the benefit that it does not disturb the operation of the tapered amplifier. The optical molasses used an intensity of $5\,$mW/cm$^2$ and a red detuning of $30\,$MHz, applied for $5\,$ms. Lower temperatures are likely with increased detunings, variable intensity and a larger grating angle $\theta$.\\ \\
\noindent
{\bf Comparison to a 6-beam MOT.}
To allow fair, direct comparison between the grating MOT and a 6-beam MOT, we used the optimal detuning of $\Delta \sim -2\Gamma$ (see Fig.~\ref{fig:det}b). All beams of a large 6-beam MOT were apertured to ensure relatively uniform intensity across an overlap volume of $1.0\,$cm$^3.$ At increasing intensities the atom number was observed to saturate at $8(1)\times 10^7$ with a single-beam intensity of $7\,$mW/cm$^2$.
\\ \\
\noindent
{\large \bf Acknowledgements}\\
We acknowledge EPSRC for support, C.N.'s Knowledge Transfer account and J.C.'s support fund. Also, the ESA through ESTEC project TEC-MME/2009/66, the CEC FP7 through project 247687 (AQUTE), the Wellcome Trust (089245/Z/09/Z), NPL’s strategic research programme, and the UK National Measurement Office. P.G.\ is supported by the Royal Society of Edinburgh and E.H.\ by the Royal Society. Chip A was fabricated by Mir Enterprises Ltd. We thank P.\ Edwards for his assistance with the SEM insets in Fig.\,2a and 2b. All other SEM images in Fig.\,2 are courtesy Kelvin Nanotechnology Ltd., who fabricated chips B-D at the James Watt Nanofabrication Centre. We also thank J.\ P.\ Griffith and G.\ A.\ C.\ Jones for assistance with GaAs e-beam lithography.\\ \\
\noindent
{\large \bf Author contributions}\\
C.N., M.V., P.G., E.R.\ and A.A.\ constructed and maintained the apparatus. C.N., J.C.\ and A.A.\
took the data which was analysed by J.C.\ and A.A. Chip A was designed by J.C.\ and E.H. Chips
B-D were designed by E.R.\ and A.A.\ with fabrication directed by P.S., A.S.\ and C.I. The manuscript was written by J.C., E.H.\ and A.A.\ with comments from all authors. \\ \\


\begin{thebibliography}{99}

\bibitem{takamoto05}
Takamoto, M., Hong, F.-L., Higashi, R.\ \& Katori,  H.\
An optical lattice clock.\
\doi{10.1038/nature03541}{\textit{Nature} \textbf{435}, 321-324 (2005)}.

\bibitem{deutsch10}
Deutsch, C.\ \textit{et al}.\ 
Spin self-rephasing and very long coherence times in a trapped atomic ensemble.\
\doi{10.1103/PhysRevLett.105.020401}{\textit{Phys.\ Rev.\ Lett}.\ \textbf{105}, 020401 (2010)}.

\bibitem{buning11}
Buning, G.~K.\ \textit{et al}.\
Extended coherence time on the clock transition of optically trapped rubidium.\
\doi{10.1103/PhysRevLett.106.240801}{\textit{Phys.\ Rev.\ Lett}.\ \textbf{106}, 240801 (2011)}.

\bibitem{bodart10}
Bodart, Q.\ \textit{et al}.\ 
A cold atom pyramidal gravimeter with a single laser beam.\
\doi{10.1063/1.3373917 }{\textit{Appl.\ Phys.\ Lett}.\ \textbf{96}, 134101 (2010)}.

\bibitem{poli11}
Poli, N.\ \textit{et al}.\
Precision measurement of gravity with cold atoms in an optical lattice and comparison with a classical gravimeter.\
\doi{10.1103/PhysRevLett.106.038501}{\textit{Phys.\ Rev.\ Lett}.\ \textbf{106}, 038501 (2011)}.

\bibitem{lamporesi08}
Lamporesi, G., Bertoldi, A., Cacciapuoti, L., Prevedelli, M.\ \& Tino, G.~M.\
Determination of the Newtonian gravitational constant using atom interferometry.\
\doi{10.1103/PhysRevLett.100.050801}{\textit{Phys.\ Rev.\ Lett}.\ \textbf{100}, 050801 (2008)}.

\bibitem{specht11}
Specht, H.~P.\ \textit{et al}.\
A single-atom quantum memory.\
\doi{10.1038/nature09997}{\textit{Nature} \textbf{473}, 190-193 (2011)}.

\bibitem{ritter12}
Ritter, S.\ \textit{et al}.\
An elementary quantum network of single atoms in optical cavities.\
\doi{10.1038/nature11023}{\textit{Nature} \textbf{484}, 195-201 (2012)}.

\bibitem{bakr10}
Bakr, W.~S.\ \textit{et al}.\ 
Probing the superfluid-to-Mott insulator transition at the single-atom Level.\
\doi{10.1126/science.1192368}{\textit{Science} \textbf{329}, 547-550 (2010)}.

\bibitem{sherson10}
Sherson, J.~F.\ \textit{et al}.\ 
Single-atom-resolved fluorescence imaging of an atomic Mott insulator.\
\doi{10.1038/nature09378}{\textit{Nature} \textbf{467}, 68-72 (2010)}.

\bibitem{weimer10}
Weimer, H., M\"uller, M., Lesanovsky, I., Zoller, P.\ \& B\"uchler, H.~P.\
A Rydberg quantum simulator.\
\doi{10.1038/NPHYS1614}{\textit{Nature Phys}.\ \textbf{6}, 382-388 (2010)}.

\bibitem{muller10}
M\"uller, H., Peters, A.\ \& Chu, S.\
A precision measurement of the gravitational redshift by the interference of matter waves.\
\doi{10.1038/nature08776}{\textit{Nature} \textbf{463}, 926-930 (2010)}.

\bibitem{knappe03}
Knappe, S.\ \textit{et al}.\ 
A microfabricated atomic clock.\
\doi{10.1063/1.1787942}{\textit{Appl.\ Phys.\ Lett}.\ \textbf{85}, 1460-1462 (2004)}; actual products at \href{www.symmetricom.com}{www.symmetricom.com}.

\bibitem{shah07}
Shah, V., Knappe, S., Schwindt, P.~D.~D.\ \& Kitching, J.\
Subpicotesla atomic magnetometry with a microfabricated vapour cell.\
\doi{10.1038/nphoton.2007.201}{\textit{Nature Photonics} \textbf{1}, 649-652 (2007).}

\bibitem{vanzoest10}
van Zoest, T.\ \textit{et al}.\
Bose-Einstein condensation in microgravity.\
\doi{10.1126/science.1189164}{\textit{Science} \textbf{328}, 1540-1543 (2010)}.

\bibitem{atomchipbook}
Reichel, J.\ \& Vuleti\'c, V., eds.\
\href{http://www.amazon.co.uk/Atom-Chips-Jakob-Reichel/dp/3527407553}{\textit{Atom Chips} (Wiley, 2011)}.

\bibitem{pollock09}
Pollock, S., Cotter, J.~P., Laliotis, A.\ \& Hinds, E.~A.\
Integrated magneto-optical traps on a chip using silicon pyramid structures.\
\doi{10.1364/OE.17.014109}{\textit{Opt.\ Express} \textbf{17}, 14109-14114 (2009)}.

\bibitem{pollock11}
Pollock, S., Cotter, J.~P., Laliotis,  A., Ramirez-Martinez, F. \& Hinds, E.~A.\
Characteristics of integrated magneto-optical traps for atom chips.\
\doi{10.1088/1367-2630/13/4/043029}{\textit{New J.\ Phys}.\ \textbf{13} 043029 (2011)}.

\bibitem{raab87}
Raab, E.~L., Prentiss, M., Cable, A., Chu, S. \& Pritchard, D.\ E.\
Trapping of neutral sodium atoms with radiation pressure.\
\doi{10.1103/PhysRevLett.59.2631}{\textit{Phys.\ Rev.\ Lett}.\ \textbf{59}, 2631-2634 (1987)}.



\bibitem{hansel01}
H\"ansel, W., Hommelhoff, P., H\"ansch, T.\ W. \& Reichel, J.\
Bose-Einstein condensation on a microelectronic chip.\
\doi{10.1038/35097032}{\textit{Nature} \textbf{413}, 498-501 (2001)}.

\bibitem{schumm05}
Schumm, T.\ \textit{et al}.\ 
Matter-wave interferometry in a double well on an atom chip.\
\doi{10.1038/nphys125}{\textit{Nat.\ Phys}.\ \textbf{1}, 57-62 (2005)}.

\bibitem{baumgartner10}
Baumg\"artner, F.\ \textit{et al}.\ 
Measuring energy differences by BEC interferometry on a chip.\
\doi{10.1103/PhysRevLett.105.243003}{\textit{Phys.\ Rev.\ Lett}.\ \textbf{105}, 243003 (2010)}.

\bibitem{riedel10}
Riedel, M.~F.\ \textit{et al}., 
Atom-chip-based generation of entanglement for quantum metrology.\
\doi{10.1038/nature08988}{\textit{Nature} \textbf{464}, 1170-1173 (2010)}.

\bibitem{coldquanta}
\href{http://www.coldquanta.com}{http://www.coldquanta.com}

\bibitem{lee96}
Lee, K.~I., Kim, J.~A., Noh, H.~R. \& Jhe, W.\
Single-beam atom trap in a pyramidal and conical hollow mirror.\
\doi{10.1364/OL.21.001177}{\textit{Opt.\ Lett}.\ \textbf{21}, 1177-1179 (1996)}.

\bibitem{vangeleyn09}
Vangeleyn, M., Griffin, P.~F., Riis, E.\ \& Arnold, A.~S.\
Single-laser, one beam, tetrahedral magneto-optical trap.\
\doi{10.1364/OE.17.013601}{\textit{Opt.\ Express} \textbf{17}, 13601-13608 (2009)}.

\bibitem{vangeleyn10}
Vangeleyn, M., Griffin, P.~F., Riis, E.\ \& Arnold, A.~S.\
Laser cooling with a single laser beam and a planar diffractor.\
\doi{10.1364/OL.35.003453}{\textit{Opt.\ Lett}.\ \textbf{35}, 3453-3455 (2010)}.

\bibitem{lind}
Lindquist, K., Stephens, M.\ \& Wieman, C.~E.\
Experimental and theoretical study of the vapor-cell Zeeman optical trap.\
\doi{10.1103/PhysRevA.46.4082}{\textit{Phys.\ Rev.\ A} \textbf{46}, 4082 (1992)}.

\bibitem{dalibard89}
Dalibard, J.\ \& Cohen-Tannoudji, C.\
Laser cooling below the Doppler limit by polarisation gradients: simple theoretical models.\
\doi{10.1364/JOSAB.6002023}{\textit{J.\ Opt.\ Soc.\ Am.\ B} \textbf{6}, 2023-2045 (1989)}.

\bibitem{blochrev}
Bloch, I., Dalibard, J.\	\& Nascimb\`{e}ne, S.\
Quantum simulations with ultracold quantum gases.\
\doi{10.1038/nphys2259}{\textit{Nature Phys}.\ \textbf{8}, 267-276 (2012)}.


\bibitem{reichel99}
Reichel, J., H\"ansel, W.\ \& H\"ansch, T.~W.\
Atomic micromanipulation with magnetic surface traps.\
\doi{10.1103/PhysRevLett.83.3398}{\textit{Phys.\ Rev.\ Lett}.\ \textbf{83}, 3398-3401 (1999)}.


\end{thebibliography}
\end{document}